\documentclass[11pt,aps,nofootinbib]{revtex4}
\usepackage{amsmath}
\usepackage{graphicx}

\def\lsim{\lesssim}

\newcommand{\Gammamat}{\boldsymbol \Gamma}
\newcommand{\Rmat}{\mathbf R}
\renewcommand{\vec}[1]{\boldsymbol #1}
\DeclareMathOperator{\trace}{tr}
\DeclareMathOperator{\Trace}{Tr}

\DeclareMathOperator{\sign}{sgn}

\begin{document}

\title{Superfluidity within Exact Renormalisation Group  approach}
\author{Boris Krippa}
\affiliation{Department of Physics, University of Surrey, Guildford
Surrey, GU2 7XH, UK}
\date{\today}
\begin{abstract}
The application of the exact renormalisation group to a 
many-fermion system with a short-range attractive force is studied. We assume a simple  ansatz for 
the effective action with effective bosons, describing pairing effects and 
 derive a set of approximate flow equations for 
the effective coupling including boson and fermionic fluctuations. 
 The  phase transition to a 
phase with broken symmetry is found at a critical value of the running 
scale.  The mean-field results are recovered  if   boson-loop effects are omitted.
The calculations with two different forms of the regulator was shown to lead to a 
 similar results. 


\end{abstract}
\maketitle

There is a growing interest in applying the Exact Renormalisation Group (ERG) formalism 
to few- and many-body systems \cite{Bir,Kr1,Kr2,Fri} when the underlying interaction is essentially non-perturbative.
Regardless of the details all ERG-based approaches share the same distinctive feature,
 a successive elimination/suppression of some modes, resulting in effective interaction between the remaining degrees of freedom \cite{Wi74}.
One specific way of implementing such a procedure is to eliminate modes by 
 applying a momentum-space blocking transformation
with some physically motivated cutoff. The effect of varying a cutoff is described by  nonlinear 
ERG evolution equations, which include the effect of the elliminated modes. By solving the ERG equations one can
 find a scale dependence of the coupling constants and thus determine a path in the space of  Lagrangean functionals.
The ERG formalism is closely linked to an other approach which has become increasingly popular in both few- and many-body physics,
Effective Field Theory (EFT). EFT is also based on a separation of scales, removal of some (mainly high-energy) degrees of freedom
and use the effective degrees of freedom instead of the fundamental ones.
 
In a sense, EFT and ERG compliment each other. EFT can provide the guidance for   fixing the
 initial conditions and ERG can be used as an analytic method to study the evolution of the system as a function of some generic 
scale parameter. This is especially useful in the case of truly non-perturbative problems, where there are no small parameters one
 can  expand in. This situation is quite common both in few- and many-body problem. One notes that, although we will focus
on the systems consisting of nucleons, many aspects of the following discussion are relevant for the other types of fermionic systems,
especially fermionic atoms in traps.

Probably the most important dynamical feature of the nucleon-nucleon interaction is the unnaturally large scattering length that makes 
a perturbative expansion meaningless both for free nucleons and in nuclear matter. In addition, the presence of the Fermi momentum 
signals the appearance of another scale which further complicates the use of any perturbative technique.
One notes that the use of the perturbation theory for the two-nucleon system in vacuum can be justified only for the
 hypothetical case of weakly 
interacting nucleons with a small scattering length.In medium, however, even a weak attraction between nucleons
 may lead to the intrinsically nonperturbative  phenomenon,  the superfluidity,  characterised  by
  rearrengement of the ground
state and appearance of the gap in the spectrum.
 The fermions form correlared pairs which, depending on the strength of the interaction, may lead  to different physical regimes.
The weak coupling regime (BCS phase) corresponds to a pair with the spatial size much larger then the radius of the interaction 
so that no actual
bound two-body subsystem is formed, while in the strong regime corresponding to the Bose-Einstein Condensation (BEC) the fermion pairs form 
compact deeply bound two-body states. 

It would be an appealing  to describe all these regimes by starting at a large initial scale  with some EFT motivated effective action 
 and then run the scale down to the physical point where the scale parameter vanishes.
 Thus the results 
 will depend on one physical parameter, the scattering length of the nucleon-nucleon interaction in free space. 

The ideal tool to treat this problem is provided by a variant of the ERG approach, based on the  average
effective  action (AEA) \cite{Wet1}.     
The (AEA) is the generating functional of the one-particle irreducible (1PI) correlation
function in the presence of an infrared cutoff scale $k$. Only fluctuations with momenta
larger then  $k$ are taken into account. For  $k \rightarrow 0$ all fluctuations are 
 included and we arrive at standard effective action from which all physical correlation 
functions can be extracted. The evolution equation for the AEA has the following one-loop form
\begin{equation}
\partial_k\Gamma=-\frac{i}{2}\,\Trace \left[(\partial_kR)\,
(\Gamma^{(2)}-R)^{-1}\right].
\end{equation}
Here $\Gamma^{(2)}$ is the second functional derivative of the AEA  taken with respect to all types of field included in the action.
and $R$ is a regulator which should suppress the contributions of states with momenta
less than or of the order of running scale $k$. To recover the full effective action we require $R(k)$ to vanish as $k\rightarrow$ 0, in other
respects its form is rather arbitrary. The concrete functional form of the regulator has no effect on physical results provided
 no approximations/truncations were made. In practice,
however, approximations/truncations are  always required to render the system of the evolution equations finite and solvable. Therefore, some
 dependence on the functional 
form of regulator is inevitable. The simplest way to estimate this dependence is to solve the system of evolution equations using several choices
 of regulators. Of course, it does not garantee the fully quantitative estimate of the errors introduced but may at least give an idea about the 
size of the corresponding uncertainty.
\section{Ansatz for $\Gamma$}\label{sec:ansatz} 
We demand that at high scale our theory  be a purely fermionic theory with the contact interaction decsribed by  the lagrangian 
\begin{equation}
{\cal L}_i=-\frac{1}{4}\,C_0\left(\psi^\dagger\sigma_2\psi^{\dagger{\rm T}}\right)
\left(\psi^{\rm T}\sigma_2\psi\right).
\end{equation}
Since we are interested in the appearance of the correlated fermion pairs in a physical 
ground state, we need to parametrise our effective action in a way that
can describe the qualitative change in the physics when this occurs.
A natural way to do this is to introduce a boson field whose vacuum 
expectation value (VEV) describes this correlated pair  \cite{wein94}.At  
the start of the RG evolution, the boson 
field is not dynamical and is introduced through a 
Hubbard-Stratonovich transformation of the four-point interaction.
As we integrate out more and more of the fermion degrees of freedom by 
running $k$ to lower values, we generate dynamical terms in the bosonic
effective action. 
 In this paper we treat a single species of fermion. The corresponding ansatz 
for the  boson-fermion effective action can be written as
\begin{eqnarray}
\Gamma[\psi,\psi^\dagger,\phi,\phi^\dagger,\mu,k]&=&\int d^4x\,
\left[\phi^\dagger(x)\left(Z_\phi\, (i \partial_t +2\mu)
+\frac{Z_m}{2m}\,\nabla^2\right)\phi(x)-U(\phi,\phi^\dagger)\nonumber\right.\\
\noalign{\vskip 5pt}
&&\qquad\qquad+\psi^\dagger\left( Z_\psi (i \partial_t+\mu)
+\frac{Z_M}{2M}\,\nabla^2\right)\psi\nonumber\\
\noalign{\vskip 5pt}
&&\qquad\qquad\left.- g \left(\frac{i}{2}\,\psi^{\rm T}\sigma_2\psi\phi^\dagger
-\frac{i}{2}\,\psi^\dagger\sigma_2\psi^{\dagger{\rm T}}\phi\right)\right].
\label{eq:ansatz}
\end{eqnarray}
Here $M$ is the mass of the fermions in vacuum and the factor $1/2m$ in the 
boson kinetic term is chosen simply to make $Z_m$ dimensionless. The
couplings, the chemical potential $\mu$,  the 
wave-function renormalisations $Z_{\phi,\psi}$ and the kinetic-mass 
renormalisations $Z_{m,M}$ all run with on $k$, the scale of the regulator. 

The bosons are , in principle coupled to the chemical potential via a quadratic term in $\phi$, 
but this can be absorbed into the potential by defining $\bar U=U-2\mu Z_\phi\phi^\dagger\phi$.

We expand this potential about its minimum, $\phi^\dagger\phi=\rho_0$, so that the coefficients $u_i$ are defined at $\rho=\rho_0$,  
\begin{equation}
\bar U(\rho)= u_0+ u_1(\rho-\rho_0)
+\frac{1}{2}\, u_2(\rho-\rho_0)^2
+\frac{1}{6}\, u_3(\rho-\rho_0)^3+\cdots,
\label{eq:potexp}
\end{equation}
where we have introduced $\rho=\phi^\dagger\phi$. The phase of the system is determined by the coefficient $u_1$. In the symmetric phase
we have $\rho_0=0$ so that the expansion takes the form 
\begin{equation}
\bar U(\rho)= u_0+u_1\rho+\frac{1}{2}\, u_2\rho^2+\cdots.
\label{eq:potexps}
\end{equation}
The potential in the condensed phase 
can be simplified to
\begin{equation}
\bar U(\rho)= u_0+\frac{1}{2}\, u_2(\rho-\rho_0)^2+\cdots.
\label{eq:potexpc}
\end{equation}
In our current work we shall truncate this potential at quartic order in the field
(order $\rho^2$). However, the fact that we define our coupling constants at the 
minimum of the potential does mean that we need to consider the next term in the
expansion. This will allow us to treat the implicit dependence of the
coefficients on $\rho_0$.
We treat the wave function renormalisation factor for the bosons
in the same way, expanding it about $\rho=\rho_0$ as
\begin{equation}
Z_\phi(\rho)= z_{\phi 0}+ z_{\phi 1}(\rho-\rho_0)+\cdots.
\label{eq:Zphiexp}
\end{equation}
 The other couplings and renormalisation
factors can be treated similarly.

The fermions are not dressed at this point and the bosons are just auxiliary fields and so we can assume that  
$Z_\psi(K)=1,\qquad Z_M(K)=1.$

We have a choice between following the evolution for fixed chemical potential,
and allowing $\mu$ to run so that the fermion density is kept fixed. The region 
corresponding to a Bose-Einstein condensate (BEC) of tightly bound pairs
corresponds to negative values of $\mu$ and is not accessible to evolution at 
fixed chemical potential. We therefore follow the evolution at fixed density, $n$.
In this context it is convenient to define the Fermi momentum, $p_F$, corresponding
to this density and the chemical potential at the starting scale by
\begin{equation}
p_F=(3 \pi^2 n)^{1/3},\qquad \mu(K)=\frac{1}{2M}\,p_F^2.
\end{equation}
This connection between $\mu$ and $p_F$ only holds in the symmetric phase, where
$\mu$ does not run.

\section{Evolution equations: general structure}
In this section we consider the general structure of the evolution equations.
We start by considering the simpler case of 
evolution at constant chemical potential.
The boson potential $\bar U$ is obtained by evaluating the effective action for
uniform boson fields. It is given by
\begin{equation}
\partial_k \bar U=-\frac{1}{{\cal V}_4}\,\partial_k\Gamma,
\end{equation}
where ${\cal V}_4$ is the volume of spacetime.
Substituting our expansion of $\bar U$, Eq.~(\ref{eq:potexp}), on the left-hand
side leads to a set of ordinary differential equations for the $u_n$.
 We choose to evolve following the minimum of the 
potential and use the expansion around the minimum $\rho=\rho_0(k)$ to define the 
$u_n$, as in Eq.~(\ref{eq:potexpc}). In the symmetric phase, $\rho=0$, it 
gives the following set of equations
\begin{equation}
\frac{du_n}{dk}=\left.\frac{\partial^n}{\partial \rho^n}
\Bigl(\partial_k \bar U\Bigr)\right|_{\rho=0},
\label{eq:evol0}
\end{equation}
 In the condensed phase we need 
to define the total derivative
\begin{equation}
\frac{d}{dk}=\partial_k+\frac{d\rho_0}{dk}\,\frac{\partial}{\partial\rho_0}.
\end{equation}
Acting on $\partial^n\bar U/\partial \rho^n$ with this (and taking the higher 
derivative term over to the LHS) gives a set of equations, each of which is coupled
to the coefficient of the next term through the evolution of $\rho_0$,
\begin{equation}
\frac{du_n}{dk}-u_{n+1}\,\frac{d\rho_0}{dk}=\left.\frac{\partial^n}{\partial \rho^n}
\Bigl(\partial_k \bar U\Bigr)\right|_{\rho=\rho_0}.
\end{equation}
The coefficient $u_1$ is special since it vanishes for the expansion around the 
minimum. Imposing the condition $u_1(k)=0$ gives an equation for the evolution
of $\rho_0$,
\begin{equation}
-u_{2}\,\frac{d\rho_0}{dk}=\left.\frac{\partial}{\partial \rho}
\Bigl(\partial_k \bar U\Bigr)\right|_{\rho=\rho_0}.
\end{equation}
The coefficient $u_0$ provides information on the energy density of matter.
It satisfies the equation 
\begin{equation}
\frac{du_0}{dk}=\partial_k \bar U|_{\rho=\rho_0},
\end{equation}
which does not couple back into the other equations. 

In the condensed phase, we could truncate our potential at quadratic order, as 
we do on the right-hand sides of these equations, and simply set $u_3=0$ on the 
left hand-side of the equation for $u_2$. However, a better approximation can be 
obtained by substituting the exact form for $u_3(k)$ taken from the evolution 
with fermion loops only, as described in below. This is the
approach we adopt here. It has the benefit of providing an approximation to 
$u_2(k)$ and $\rho_0(k)$ that becomes exact in situations where boson 
loops can be neglected.

 The evolution of the boson wave-function renormalisation factor  $Z_\phi$ can be obtained from
\begin{equation}
\partial_k Z_\phi=\frac{1}{{\cal V}_4}\left.\frac{\partial}{\partial p_0}
\left(\frac{\partial^2}{\partial\eta\partial\eta^\dagger}\,\partial_k\Gamma
\right)_{\eta=0}\,\right|_{p_0=0}.
\end{equation}
If we substitute our expansion, Eq.~(\ref{eq:Zphiexp}), we get another set of 
coupled equations in the condensed phase. Only the first of these is of interest
within our current truncation
\begin{equation}
\frac{dz_{\phi 0}}{dk}-z_{\phi 1}\,\frac{d\rho_0}{dk}
=\frac{1}{{\cal V}_4}\left.\frac{\partial}{\partial p_0}\left(
\frac{\partial^2}{\partial\eta\partial\eta^\dagger}\,\partial_k\Gamma
\right)_{\eta=0}\,\right|_{p_0=0,\,\rho=\rho_0}.
\end{equation}
Again, $z_{\phi 1}$ corresponds to a term beyond our current level of truncation
and so we will take the result from fermion loops only.
The evolution of $Z_\phi$ can also be deduced from
\begin{equation}
\partial_k Z_\phi=-\,\frac{1}{2}\frac{\partial^2}{\partial \mu\partial\rho}
\Bigl(\partial_k \bar U\Bigr),
\end{equation}
which gives
\begin{equation}
\frac{dz_{\phi 0}}{dk}-z_{\phi 1}\,\frac{d\rho_0}{dk}
=-\,\frac{1}{2}\left.\frac{\partial^2}{\partial \mu\partial\rho}
\Bigl(\partial_k \bar U\Bigr)\right|_{\rho=\rho_0}.
\label{eq:Zphialt}
\end{equation}
The evolution equations for the other couplings ($Z_m, Z_M, Z_\psi, Z_g$) can be derived in
 a similar manner. However, in this paper we allow to run only $Z_\phi$, parameters in the potential and 
chemical potential since this is the minimal set needed to include the effective boson dynamics and 
study the BCS-BEC crossover.

The fermion number density is given by 
\begin{equation}
n=-\,\left.\frac{\partial \bar U}{\partial \mu}\right|_{\rho=\rho_0}.
\end{equation}
The evolution equation for 
$n$ can be written as
\begin{equation}
\frac{dn}{dk}-2z_{\phi 0}\,\frac{d\rho_0}{dk}
=-\left.\frac{\partial}{\partial \mu}
\Bigl(\partial_k \bar U\Bigr)\right|_{\rho=\rho_0}.
\end{equation}

The equations constructed so far describe the evolution at constant $\mu$. 
If we want to follow the evolution at constant density we must allow $\mu$ to 
run with $k$. In this case we define the total derivative
\begin{equation}
\frac{d}{dk}=\partial_k+\frac{d\rho_0}{dk}\,\frac{\partial}{\partial\rho_0}
+\frac{d\mu}{dk}\,\frac{\partial}{\partial\mu}.
\end{equation}
Applying this to $\partial\bar U/\partial \mu$ at $\rho=\rho_0$ gives the
evolution equation for $n$
\begin{equation}
\frac{dn}{dk}-2z_{\phi 0}\,\frac{d\rho_0}{dk}+\chi\,\frac{d\mu}{dk}
=-\left.\frac{\partial}{\partial \mu}
\Bigl(\partial_k \bar U\Bigr)\right|_{\rho=\rho_0},
\end{equation}
where we have introduced the fermion-number susceptibility
\begin{equation}
\chi=\left.\frac{\partial^2\bar U}{\partial \mu^2}
\right|_{\rho=\rho_0}.
\end{equation}
If $n$ is kept constant ($dn/dk=0$) this becomes
\begin{equation}
-2z_{\phi 0}\,\frac{d\rho_0}{dk}+\chi\,\frac{d\mu}{dk}
=-\left.\frac{\partial}{\partial \mu}
\Bigl(\partial_k \bar U\Bigr)\right|_{\rho=\rho_0}.
\label{eq:muevol}
\end{equation}
This equation describes evolution of chemical potential in the broken phase

 The remaining set of evolution equations to be solved (in the broken phase) is
\begin{eqnarray}
\frac{du_0}{dk}+n\,\frac{d\mu}{dk}
&=&\left.\partial_k \bar U\right|_{\rho=\rho_0},\\
\noalign{\vskip 5pt}
-u_2\,\frac{d\rho_0}{dk}+2z_{\phi 0}\,\frac{d\mu}{dk}
&=&\left.\frac{\partial}{\partial \rho}
\Bigl(\partial_k \bar U\Bigr)\right|_{\rho=\rho_0},\\
\noalign{\vskip 5pt}
\frac{du_2}{dk}-u_3\,\frac{d\rho_0}{dk}+2z_{\phi 1}\,\frac{d\mu}{dk}
&=&\left.\frac{\partial^2}{\partial \rho^2}
\Bigl(\partial_k \bar U\Bigr)\right|_{\rho=\rho_0},\\
\noalign{\vskip 5pt}
\frac{dz_{\phi 0}}{dk}-z_{\phi 1}\,\frac{d\rho_0}{dk}+\frac{1}{2}\,\chi'\,
\frac{d\mu}{dk}
&=&-\,\frac{1}{2}\left.\frac{\partial^2}{\partial \mu\partial\rho}
\Bigl(\partial_k \bar U\Bigr)\right|_{\rho=\rho_0},\\
\noalign{\vskip 5pt}
\frac{dz_{m0}}{dk}-z_{m1}\,\frac{d\rho_0}{dk}+\alpha_m\,\frac{d\mu}{dk} 
&=&-\,\frac{1}{{\cal V}_4}\left.\frac{\partial}{\partial (p^2)}\left(
\frac{\partial^2}{\partial\eta\partial\eta^\dagger}\,\partial_k\Gamma
\right)_{\eta=0}\,\right|_{p^2=0,\,\rho=\rho_0}.
\end{eqnarray}
where we have defined
\begin{equation}
\chi'=\left.\frac{\partial^3\bar U}{\partial \mu^2\partial\rho}
\right|_{\rho=\rho_0}, \qquad \alpha_m=\frac{1}{{\cal V}_4}\left.
\frac{\partial^2}{\partial\mu\partial (p^2)}\left(
\frac{\partial^2}{\partial\eta\partial\eta^\dagger}\,\partial_k\Gamma
\right)_{\eta=0}\,\right|_{p^2=0,\,\rho=\rho_0}
\end{equation}
The set of evolution equations in symmetric phase can easily be recoved 
using the fact that chemical potential does not run in symmetric phase and that
$\rho_0 =0$.
The left-hand sides of these equations  contain a number of coefficients that lie beyond 
our current level of truncation, such as $\chi$, $u_3$ and $z_{\phi 1}$. 
 We propose to replace these by their exact expressions obtained 
from evolution with fermion loops only. The formal derivation of the corresponding expressions will 
be considered in more details below.

\section{Choice of cut-off}

In the bosonic sector, we take the regulator to be an additional
quadratic term, proportional to $\phi^\dagger(x)\phi(x')$. In the 
representation used to write down the second derivatives above, 
it has the matrix structure
\begin{equation}
\Rmat_B(q,k)=\left(\begin{array}{cc} R_B(q,k) & 0 \cr 0 & R_B(q,k) \end{array}
\right), \qquad R_B(q,k)=\frac{k^2}{2m}\,f(q/k),
\end{equation}

where $f(x)\rightarrow 1$ as $x\rightarrow 0$.

In the fermion case, our regulator should be positive for particle states 
($Z_M q^2/2M>Z_\psi\mu$) and negative for hole states ($Z_M q^2/2M<Z_\psi\mu$). 
It should suppress 
the contributions of states with energies near $\mu$. One easy way to
ensure this would be to use the off-diagonal  regulator and so generates an artificial gap
in the fermion spectrum around $\mu$. However such a regulator could not 
be used without a Fermi sea and so would not allow us to connect our results 
in matter to the interaction between the fermions in vacuum. We therefore
choose our regulator to have the structure 
\begin{equation}
\Rmat_F(q,p_F,k)=\left(\begin{array}{cc} \sign
(q-p_\mu)\,R_F(q,p_F,k)&0\cr 
\noalign{\vskip 5pt}
0&-\sign(q-p_\mu)\,R_F(q,p_F,k)\end{array}
\right),
\end{equation}
where we have introduced
\begin{equation}
p_\mu=\sqrt{\frac{Z_\psi 2M\mu}{Z_M}},
\end{equation}
the Fermi momentum corresponding to the (running) value of $\mu$.

The function $R_F(q,p_F,k)$ should suppress the contributions 
of states with momenta near the Fermi surface, $|q-p_F|\lsim k$. Once a 
a large gap has appeared in the fermion spectrum, there are no low-energy fermion
excitations and so the fermionic regulator plays little further role. However, 
while the gap is zero or small, it is crucial that the sign of the regulator
match that of the energy, $Z_M q^2/2M-Z_\psi\mu$, and hence it is $\mu$ which
appears in the sign functions.

In order for to be sure that we are matching onto the same bare NN 
interaction at the starting scale $K$, the fermionic regulator should 
satisfy $R_F(q,p_F,K)\simeq R_F(q,0,K)$.

\section{Driving terms: potential}

In this section we derive the evolution equations for the parameters of the potential. Calculating the fermion propagator, 
multiplying by $\partial_k\Rmat_F$ and taking the matrix trace gives
\begin{equation}
\frac{1}{2}\,\trace\left[(\partial_k\Rmat_F)\,
(\Gammamat ^{(2)}_{FF}-\Rmat_F)^{-1}\right]
=\frac{2E_{FR}(q,p_F,k)\,\sign(q-p_\mu)\,\partial_k R_F(q,p_F,k)}
{Z_\psi^2 q_0^2-E_{FR}(q,p_F,k)^2-\Delta^2+i\epsilon}.
\end{equation}
where
\begin{equation}
E_{FR}(q,p_F,k)=\frac{Z_M}{2 M}\,q^2-Z_\psi\mu+R_F(q,p_F,k)\,\sign(q-p_\mu),  \qquad \Delta^2=g^2\phi^\dagger\phi.
\label{eq:defEFR}
\end{equation}

The poles in this propagator occur at
\begin{equation}
q_0=\pm\frac{1}{Z_\psi}\sqrt{E_{FR}(q,p_F,k)^2+\Delta^2}.
\end{equation}
At $k=0$ ($R_F=0$) in the condensed phase, these become
\begin{equation}
q_0=\pm\frac{1}{Z_\psi}\sqrt{\left(\frac{Z_M}{2M}(q^2-p_F^2)
\right)^2+\Delta^2},
\end{equation}
and so the gap in the fermion spectrum at $q=p_F$ is $2\Delta/Z_\psi$.

The corresponding boson matrix trace can be worked out in a similar way. Putting everything together gives
\begin{eqnarray}
\partial_k \bar U=-\frac{1}{{\cal V}_4}\,\partial_k\Gamma
&=&-\,\frac{1}{Z_\psi}\int\frac{d^3{\vec q}}{(2\pi)^3}\,\frac{E_{FR}}
{\sqrt{E_{FR}^2+\Delta^2}}\,\sign(q-p_\mu)\,\partial_kR_F\nonumber\\
\noalign{\vskip 5pt}
&&+\,\frac{1}{2Z_\phi}\int\frac{d^3{\vec q}}{(2\pi)^3}\,
\frac{E_{BR}}{\sqrt{E_{BR}^2-V_B^2}}
\,\partial_kR_B.\label{eq:potevol}
\end{eqnarray}
where
\begin{equation}
E_{BR}(q,k)=\frac{Z_m}{2m}\,q^2+u_1
+u_2(2\phi^\dagger\phi-\rho_0)+R_B(q,k), \qquad V_B= u_2\phi^\dagger\phi.
\end{equation}
The driving terms in the evolution equations for the coefficients in our 
expansion, Eq.(\ref{eq:potexp}), are obtained from the derivatives of
$\partial_k \bar U$ with respect to $\rho=\phi^\dagger\phi$ so that we get in the
condensed phase 
\begin{eqnarray}
\left.\frac{\partial}{\partial \rho}
\Bigl(\partial_k \bar U\Bigr)\right|_{\rho=\rho_0}
&=&\frac{g^2}{2Z_\psi}\int\frac{d^3{\vec q}}{(2\pi)^3}\,
\frac{E_{FR}}{\Bigl(E_{FR}^2+\Delta^{(c)2}\Bigr)^{3/2}}
\,\sign(q-p_\mu)\,\partial_kR_F\nonumber\\
\noalign{\vskip 5pt}
&&+\,\frac{u_2V_B^{(c)}}{2Z_\phi}\int\frac{d^3{\vec q}}{(2\pi)^3}\,
\frac{E_{BR}^{(c)}-2V_B^{(c)}}{\Bigl(E_{BR}^{(c)2}-V_B^{(c)2}\Bigr)^{3/2}}\,
\partial_kR_B,
\label{eq:rho0broken}
\end{eqnarray}
where
\begin{equation}
E_{BR}^{(c)}(q)=\frac{Z_m}{2m}\,q^2
+u_2\rho_0+R_B(q,k);V_B^{(c)}=u_2\rho_0, \qquad \Delta^{(c)}=g\sqrt{\rho_0},\\
\end{equation}
and similarly for the second derivative of
$\partial_k \bar U$ with respect to $\rho=\phi^\dagger\phi$.
The equations for the couplings $u_{1(2)}$ in the both phases can  be obtained in a similar way
from the driving terms remembering that in the symmetric phase  ($\rho=\rho_0, u_1\neq 0$).
The driving term for the evolution of the fermion number density is given by
the derivative of $\bar U$ with respect to $\mu$.
The sign functions in the fermion part of Eq.~(\ref{eq:potevol}) depend on $\mu$
and so, in principle, differentiating with respect to $\mu$ could generate
surface terms. However the sign change occurs at precisely the point where
$E_{FR}$ vanishes and hence such terms do not arise. The resulting driving 
term vanishes in the symmetric phase, and so evolution at constant $n$ and
at constant $\mu$ are the same there.

For the evolution of $\mu$ in the condensed phase we get
\begin{eqnarray}
-\left.\frac{\partial}{\partial \mu}
\Bigl(\partial_k \bar U\Bigr)\right|_{\rho=\rho_0}
&=&-\int\frac{d^3{\vec q}}{(2\pi)^3}\,\frac{\Delta^{(c)2}}
{\Bigl(E_{FR}^2+\Delta^{(c)2}\Bigr)^{3/2}}\,\sign(q-p_\mu)\,\partial_kR_F\nonumber\\
\noalign{\vskip 5pt}
&&-\int\frac{d^3{\vec q}}{(2\pi)^3}\,
\frac{V_B^{(c)2}}{\Bigl(E_{BR}^{(c)2}-V_B^{(c)2}\Bigr)^{3/2}}
\,\partial_kR_B.
\end{eqnarray}

\section{Driving terms: $Z_\phi$}

For the wave function renormalisation factor, $Z_\phi$, we need to consider
a time-dependent background field. The evolution 
of $Z_\phi$ is then driven by 
\begin{equation}
\frac{1}{{\cal V}_4}\left.\frac{\partial}{\partial p_0}
\left(\frac{\partial^2}{\partial\eta\partial\eta^\dagger}\,\partial_k\Gamma
\right)_{\eta=0}\,\right|_{p_0=0}.
\end{equation}

Defining 
\begin{equation}
\Gammamat ^{(3)}_{BB\phi}
=\frac{\partial}{\partial\phi}\,\Gammamat ^{(2)}_{BB}
=\left(\begin{array}{cc} -2u_2\phi^\dagger&-2u_2\phi\cr
0&-2u_2\phi^\dagger\end{array}\right),
\end{equation}
and
\begin{equation}
\Gammamat ^{(3)}_{FF\phi}
=\frac{\partial}{\partial\phi}\,\Gammamat ^{(2)}_{FF}
=\left(\begin{array}{cc} 0&ig\sigma_2\cr
0&0\end{array}\right),
\end{equation}
we can write the relevant part of the evolution equation in the form
\begin{eqnarray}
\left.\frac{\partial^2}{\partial\eta\partial\eta^\dagger}\,\partial_k\Gamma
\right|_{\eta=0}&=&+i\,\Trace \left[(\partial_k\Rmat_F)\,
(\Gammamat ^{(2)}_{FF}-\Rmat_F)^{-1}
\,\Gammamat ^{(3)\dagger}_{FF\phi}\,
(\Gammamat ^{(2)}_{FF}-\Rmat_F)^{-1}
\,\Gammamat ^{(3)}_{FF\phi}\,
(\Gammamat ^{(2)}_{FF}-\Rmat_F)^{-1}
\right]\nonumber\\
\noalign{\vskip 5pt}
&&-i\,\Trace \left[(\partial_k\Rmat_B)\,
(\Gammamat ^{(2)}_{BB}-\Rmat_B)^{-1}
\,\Gammamat ^{(3)\dagger}_{BB\phi}\,
(\Gammamat ^{(2)}_{BB}-\Rmat_B)^{-1}
\,\Gammamat ^{(3)}_{BB\phi}\,
(\Gammamat ^{(2)}_{BB}-\Rmat_B)^{-1}\right].\nonumber\\
\end{eqnarray}
After lengthly algebra one can obtain in broken phase
\begin{eqnarray}
\frac{1}{{\cal V}_4}\left.\frac{\partial}{\partial p_0}
\left(\frac{\partial^2}{\partial\eta\partial\eta^\dagger}\,\partial_k\Gamma
\right)_{\eta=0}\,\right|_{p_0=0}
&=&-\frac{g^2}{4}\int\frac{d^3{\vec q}}{(2\pi)^3}\,
\frac{2E_{FR}^2-\Delta^{(c)2}}
{\Bigl(E_{FR}^2+\Delta^{(c)2}\Bigr)^{5/2}}\,\sign(q-p_\mu)\,\partial_kR_F
\nonumber\\
\noalign{\vskip 5pt}
&&-\frac{u_2V_B^{(c)}}{2}\int\frac{d^3{\vec q}}{(2\pi)^3}\,
\frac{2E_{BR}^{(c)2}-6E_{BR}^{(c)}V_B^{(c)}+V_B^{(c)2}}
{\Bigl(E_{BR}^{(c)2}-V_B^{(c)2}\Bigr)^{5/2}}\,
\partial_kR_B.\nonumber\\
\label{eq:Zphibroken}
\end{eqnarray}
The corresponding expression in symmetric phase follows rather trvially.

As a check on this result, we note that $u_1$ contains a piece $-2\mu Z_\phi$.
Hence we can also obtain the evolution of $Z_\phi$ from
\begin{equation}
-\,\frac{1}{2}\left.\frac{\partial^2}{\partial \mu\partial\rho}
\Bigl(\partial_k \bar U\Bigr)\right|_{\rho=\rho_0}.
\end{equation}
Taking the partial derivative with respect to $\mu$, as discussed at the 
end of the previous section, the result we obtain  
agrees with Eq.~(\ref{eq:Zphibroken}).

\section{Fermions only}
In order to estimate the effects of the higher order coefficients ($u_3$ etc) we use the 
ERG equations when boson loops are neglected. In this case the expressions simplify considerably and 
the total effective potential can be calculated analytically. All the needed coefficients can then
 be expracted by simple differentiation. The RG equation for the effective potential becomes  
\begin{equation}
\partial_k \bar U=-\int\frac{d^3{\vec q}}{(2\pi)^3}\,\frac{E_{FR}}
{\sqrt{E_{FR}^2+\Delta^2}}\,\sign(q-p_\mu)\,\partial_kR_F.
\end{equation}
(In the absence of boson fluctations, we have $Z_\psi=1$.)
This can be rewritten in the form
\begin{equation}
\partial_k \bar U=-\partial_k\int\frac{d^3{\vec q}}{(2\pi)^3}
{\sqrt{E_{FR}^2+\Delta^2}}\,,
\end{equation}
and immediately integrated to give
\begin{equation}
\bar U(\rho,\mu,k)=\bar U(\rho,\mu,K)-\int\frac{d^3{\vec q}}{(2\pi)^3}
\left[{\sqrt{E_{FR}(q,p_F,k)^2+\Delta^2}}
-{\sqrt{E_{FR}(q,p_F,K)^2+\Delta^2}}\right]\,.
\end{equation}
We have made explicit the dependence of the potential on the chemical
potential $\mu$.

At our starting scale $K$, we take the potential to have the form
\begin{equation}
\bar U(\rho,\mu,K)=u_0(K)+u_1(K)\,\rho.
\end{equation}
The renormalised value of $u_1(K)$can be deduced from the scattering
length. To determine $u_0(K)$, we use the fact
that at $\rho=0$ the physical potential is just that of a free Fermi gas,
measured relative to the chemical potential:
\begin{equation}
\bar U(0,\mu,0)=2\int\frac{d^3{\vec q}}{(2\pi)^3}\,E_{FR}(q,p_F,0)\theta(p_F-q),
\end{equation}
and hence 
\begin{equation}
u_0(K)=\int\frac{d^3{\vec q}}{(2\pi)^3}\,E_{FR}(q,p_F,0)
-\int\frac{d^3{\vec q}}{(2\pi)^3}\,E_{FR}(q,p_F,K)\sign(q-p_F).
\end{equation}

The physical potential (at $k=0$) is then given by
\begin{eqnarray}
\bar U(\rho,\mu,0)&=&\int\frac{d^3{\vec q}}{(2\pi)^3}\,
\left[E_{FR}(q,p_F,0)-\sqrt{E_{FR}(q,p_F,0)^2+\Delta^2}\right]\nonumber\\
\noalign{\vskip 5pt}
&&-\,\frac{M\Delta^2}{4\pi a}+\frac{\Delta^2}{2}
\int\frac{d^3{\vec q}}{(2\pi)^3}\,\frac{1}{E_{FR}(q,0,0)}\,.
\end{eqnarray}

Differentiating with respect to $\rho$ and setting the derivative equal to
zero, we find that $\Delta^2$ at the minimum satisfies
\begin{equation}
-\,\frac{M}{4\pi a}+\frac{1}{2}\int\frac{d^3{\vec q}}{(2\pi)^3}\,
\left[\frac{1}{E_{FR}(q,0,0)}-\frac{1}{\sqrt{E_{FR}(q,p_F,0)^2+\Delta^2}}
\right]=0.
\end{equation}
This is exactly the gap equation used in Ref.\cite{PB99}.

To get the number density of fermions, we can differentiate $\bar U(\rho,\mu,0)$
with respect to $\mu$. This gives 
\begin{equation}
n=\int\frac{d^3{\vec q}}{(2\pi)^3}\,\left[1
-\frac{E_{FR}(q,p_F,0)}{\sqrt{E_{FR}(q,p_F,0)^2+\Delta^2}}\right],
\end{equation}
in agreement with Ref.~\cite{PB99}.

\section{Initial conditions}

We start our evolution at some large, fixed value for the cut-off scale, $K$,
and evolve down to $k=0$. This clearly requires some initial conditions on the
parameters in our effective action. Some constraints on these have already been
mentioned in Sec.~\ref{sec:ansatz}, but we now turn to fixing specific values
for them. These should be derived from the known interactions between the 
particles in vacuum. In  the vacuum case, it is relatively 
straightforward to fix the initial values for 
the parameters at the starting scale $K$.

In matter, the detemination of our renormalised parameters is complicated by 
the fact that our fermionic cut-off depends on the Fermi momentum. The values 
of parameters such as $u_1(p_F,K)$ must thus depend on $p_F$ as well as $K$. 
Note that we assume that our starting scale $K$ is sufficiently large that
all physical effects of the Fermi sea have been completely suppressed by
our cut-off. Hence the dependence of the initial parameters on $p_F$ is
merely to compensate for the $p_F$ dependence of our regularisation
and renormalisation procedures. For example, in the case of a simple 
sharp cut-off, the maximum momentum included is $K+p_F$ in matter instead 
of $K$. It is thus particularly important to correctly renormalise 
$u_1(p_F,K)$, since this quantity cancels a linear divergence and so
can be shifted by a finite amount even for $K\rightarrow\infty$.

One could introduce a cut-off function that tends to a $p_F$-independent form
for $K\gg p_F$. However in practice a modification of the renormalisation
procedure is more convenient. In the region $K\gg p_F$, we can ignore boson
loops. The evolution of
quantities such as $u_1(p_F,K)$, $u_2(p_F,K)$, $Z_\phi(p_F,K)$ and
$Z_m(p_F,K)$ is thus similar to 
the vacuum case, except for the different cut-off. This allow us to define
$u_1(p_F,K)$ to be
\begin{equation}
\frac{u_1(p_F,K)}{g^2}=-\,\frac{M}{4\pi a}
+\frac{1}{2}\,\int\frac{d^3{\vec q}}{(2\pi)^3}\,
\left[\frac{1}{E_{FR}(q,0,0)}-\frac{\sign(q-p_F)}{E_{FR}(q,p_F,K)}\right]\, .
\label{eq:u1Kfull}
\end{equation}
This expression can be thought of as being generated by the vacuum evolution 
using a modified cut-off that interpolates smoothly between $R_F(q,p_F,k)$
for $k\gg p_F$ and $R_F(q,0,k)$ for $k\lsim p_F$. It ensures that our 
renormalised parameter $u_1(p_F,K)$, defined using $R_F(q,p_F,k)$ for large 
$k$, corresponds to the physical scattering length in vacuum.

The initial values for $u_2(p_F,K)$, $Z_\phi(p_F,K)$ and $Z_m(p_F,K)$ 
can be determined using similar procedures, although this is not so 
crucial since these quantities do not contain linearly divergent pieces 
and so all their $p_F$-dependence is suppressed by powers of $p_F/K$.
One convenient choice is to take their starting values to be zero at some
large but finite scale $K$. An alternative is to require that they tend 
to zero as $K\rightarrow\infty$.

The initial condition for the energy density is most conveniently
expressed in terms of $\tilde u_0$ which,  in the symmetric phase, is
simply given by the energy of a free Fermi gas, measured relative to the 
chemical potential, and so its initial value 
is just
\begin{equation}
\tilde u_0(K)=2\int\frac{d^3{\vec q}}{(2\pi)^3}\,E_{FR}(q,p_F,0)\theta(p_F-q).
\end{equation}

\section{Results}

We solve the evolution equations numerically with two types of cutoffs.
First, we use the smoothed step-function type of regulator (called hereafter as $R_1$)
\begin{equation}
R_{1F}=\frac{k^2}{2M}\theta_{1}(q - p_{F},k,\sigma),  \qquad R_{1B}=\frac{k^2}{2m}\theta_{1}(q,k,\sigma)
\end{equation}

where
\begin{equation}
\theta_{1}(q,k,\sigma)=\frac{1}{2 {\bf Erf}(1/\sigma)}\left[{\bf Erf}(\frac{q +k}{k \sigma})
 + {\bf Erf}(\frac{q - k}{k \sigma})\right]
\end{equation}
with $\sigma$ being a parameter determining the sharpness of the step.

Second, we use a sharp cutoff function denoted $R_2$  chosen 
 to make the calculations as simple as possible
\begin{equation}
R_{2F}=\frac{k^2}{2M}\left[((k + p_\mu)^{2}- q^2)\theta(p_{\mu}+k -q) + ((k + p_\mu)^{2}+ q^2 -2p^{2}_\mu)\theta(q - p_{\mu}+k)\right],
\end{equation}
\begin{equation}
R_{2B}=\frac{k^2}{2m}(k^2 - q^2)\theta(k -q). 
\end{equation}
Similar boson regulator was used in Ref. \cite{Bla} (see also Ref.\cite{Litim}).  

As we can see the fermion sharp cutoff consists of two terms which result in modification of the particle  and hole propagators
 respectively. 
The hole term is further modified to suppress the contribution from the surface terms, which may bring in the dangerous dependence
 of the 
regulator on the cutoff scale even at the vanishingly small $k$. As an example, we focus on the parameters relevant to neutron matter:
$M=4.76 fm^{-1}, p_{F}=1.37 fm^{-1}$. 

We first discuss the results obtained with the smooth cutoff $R_1$. We find that the value of the
physical gap  is  practically independent of either the values of the width parameter $\sigma$ (varied within some range) or
the starting scale $K$ provided $K > 5 fm^{-1}$. The results of the calculations are shown on Fig. 1.
\begin{figure}
\begin{center}
\includegraphics[width=9cm,  keepaspectratio,clip]{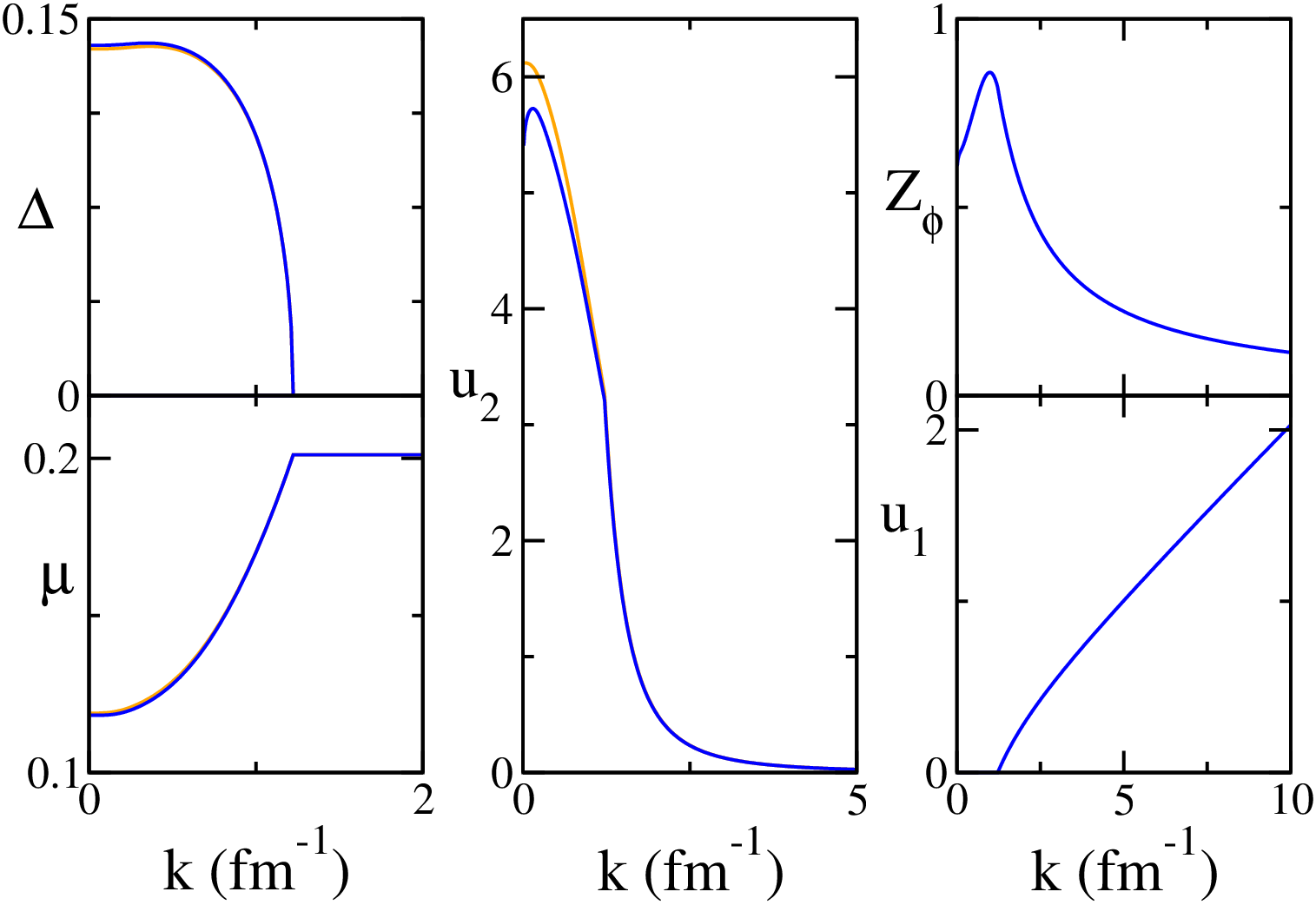}
\end{center}
\caption{\label{fig:running}Numerical solutions to the evolution equations
for infinite $a_0$ and $p_F=1.37$~fm, starting from $K=16$~fm$^{-1}$.
We show the evolution of all relevant 
parameters for the cases of fermion loops only (thin lines), and  
of bosonic loops with a running $Z_\phi$ (thick lines). All 
quantities are expressed in appropriate powers of fm$^{-1}$.}
\end{figure}
 At starting scale
 the system is in the symmetric phase and remains in this phase until $u_1$ hits zero at $k_{crit}\simeq 1.2 fm^{-1}$ where the artificial second
order phase transition to a broken phase occurs and the energy gap is formed. Already at $k \simeq 0.5$ the running scale has essentually no effect on
 the gap. 
  We found very small (on the level of 1$\%$)
 contribution to the gap from the boson loops, due to cancellations between the direct contributions to the running of the gap and indirect ones via $u_2$.
 The boson loops play much more important role  in  the evolution of $u_2$ and $Z_\phi$. In fact, they drive
both couplings to zero at $k\rightarrow 0$ although at rather slow pace. We note however, that the effect of the boson loops for the
 gap may still  be more visible if the evolution of
 the other couplings is included. The results obtained for the gap correspond to the case of the infinite negative scattering length.
 To study the BCS-BEC crossover we have to solve the evolution equation for a wider range of the scattering lengths, including the positive values.
The corresponding results for the evolution of chemical potential as a function of the parameter $p_{F}a$ are  shown in Fig.2.
\begin{figure}
\begin{center}
\includegraphics[width=6cm,  keepaspectratio,clip]{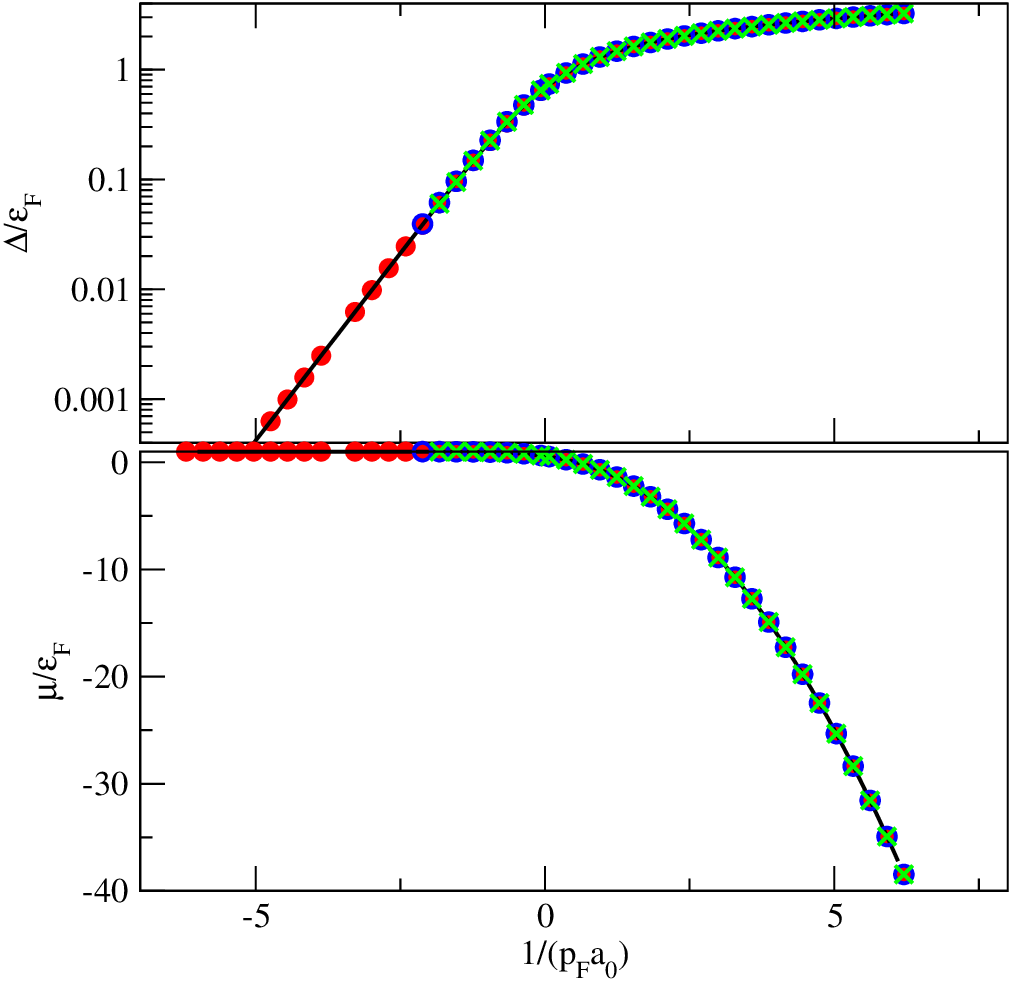}
\end{center}
\caption{\label{fig:running1}Evolution of the chemical potential}
\end{figure}
While the vacuum scattering length is large and negative, the system is in the BCS-like phase with positive chemical potential, whereas if the 
 scattering length is chosen to be large and positive  reflecting the existence of a bound state near threshold
the system ends up being the collection of weakly overlaping tightly bound pairs with negative chemical potential.

Now we turn to the results obtained with sharp cutoff (Fig.3). One immediate observation is that that the results become starting scale independent
as long as $K > 5 fm^{-1}$ similar to the smooth regulator case. However, the  phase transition occurs at lower values of the running scale
$k \simeq 0.7 fm^{-1}$. At approximately $k \simeq 0.2 fm^{-1}$ the value of the gap becomes scale independent. The
 gap evolutions obtained with the smooth and sharp regulators, being rather different at intermediate scales, approach each other with decreasing scale 
resulting in similar values for the physical gap. 
This is an encouraging results taking into account that, although the exact results must be independent of the choice of the regulator, in practice
 it is not garanteed. The same conclusion also holds for other quantities. The couplings $Z_{\phi}$ and $u_2$ first grow with scale and 
then start decreasing eventually coming to zero. Chemical potential begins to decrease at the point of phase transition and becomes scale independent at
$k\simeq 0.2 fm^{-1}$. However, in this case the numerical values of the chemical potentials obtained with different regulators differ
 by approximately $20\%$ so that this quantity is more sensitive to the details of effective action and to the trancations made. We note that the sharp 
regulator can also describe the BCS-BEC crossover, although gives somewhat smaller (negative) values of chemical potential at $p_{F}a > 1$.
\begin{figure}
\begin{center}
\includegraphics[width=9cm,  keepaspectratio,clip]{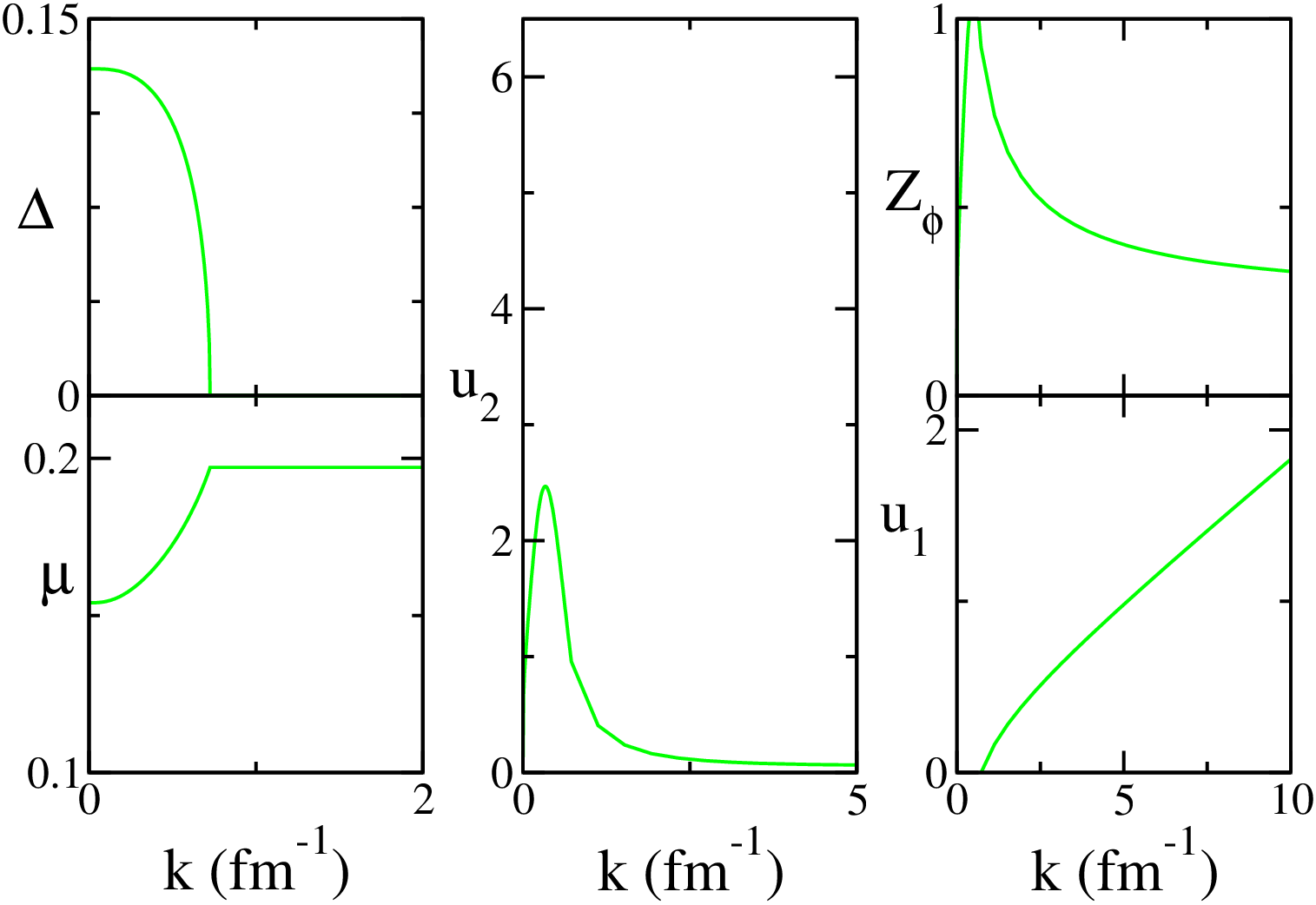}
\end{center}
\caption{\label{fig:running2}Evolution of the parameters when the sharp cut-off is used}
\end{figure}
Applying our approach to neutron  matter we find a gap 
 comparable to $\epsilon_F$, of the order of 30~MeV.
There is  a simple explanation for the smaller values (see \cite{Fayans}, \cite{Khodel} or \cite{EHJ98}).
 The argument can be given most succinctly 
for weak coupling, where the gap satisfies
\begin{equation}
\Delta=({8}/{e^2})\epsilon_F \exp\left(-\,({\pi}/{2})
\cot\bigl(\delta(p_F)\bigr) \right).
\end{equation}
For nucleon-nucleon scattering, $\cot\delta$ increases relatively quickly
with momentum and the resulting reduction in the gap is substantial.
We therefore expect that an extension of our approach to include the 
effective range should capture this physics. Indeed, if the 
``in-medium'' scattering length is identified with the Bethe-Goldstone
$G$ matrix calculated at zero energy but finite momenta \cite {Kri} then we obtain the gap $\simeq$ 8~MeV which is already compatible with the commonly 
accepted value.
Of course, this is only a crude estimate and the proper calculations should be done using the momentum dependent part of the four-fermion interaction.

There are several points where our approach could be improved. We have already mentioned above the effective range effects. We should also include running
 of all 
the couplings and treat explicitly the 
particle-hole channels (RPA phonons) since these contain important physics.
They will  allow us to remove the ``Fierz ambiguity'' associated with
our bosonisation of the underlying contact interaction \cite{JW03}. We would like to include the three-body force effects, which are required to 
satisfy the reparametrisation invariance theorem \cite{BKWM} and possibly the long-range forces.
As to further applications we plan to explore the superfluidity of the cold fermionic atoms in traps, temperature dependence of the BEC-BSC transitions and 
the formal relations of the ERG with the other many-body approaches.

It is a pleasure to thank Mike Birse, Niels Walet and Judith McGovern for very useful discussions. 

This research was funded by the EPSRC.

\end{document}